# Nonmagnetic Impurity Effect of the $S$=1/2 Spin Ladder System (pipdH)$_2$Cu$_{1-x}$Zn$_x$Br$_4$

Chiori Yokoyama, Eiichi Matsuoka and Hitoshi Sugawara

*Graduate School of Science, Kobe University, Kobe 657-8501, Japan*

Takahiro Sakurai

*Center for Supports to Research and Education Activities, Kobe University, Kobe 657-8501, Japan*

Weimin Zhang, Susumu Okubo and Hitoshi Ohta

*Molecular Photoscience Research Center, Kobe University, Kobe 657-8501, Japan*

Hikomitsu Kikuchi

*Department of Applied Physics, University of Fukui, Fukui 910-8507, Japan*

We report the synthesis and the magnetic susceptibility of (pipdH)$_2$Cu$_{1-x}$Zn$_x$Br$_4$ (pipd=piperidinium), which is a nonmagnetic impurity-doped $S$=1/2 spin ladder system. The samples were synthesized from a solution by using a slow evaporation method. Samples were confirmed to be in a single phase and to have the same crystal structure as the pure system (pipdH)$_2$CuBr$_4$ by using X-ray diffraction measurements. To check the magnetic properties, we performed magnetic susceptibility and magnetization measurements with a superconducting quantum interference device (SQUID) magnetometer. A plot of the inverse magnetic susceptibility indicates the presence of dominant antiferromagnetic coupling. The magnetic susceptibility shows a broad maximum due to low dimensionality and a spin gap behavior related to the two-leg spin ladder at low temperature. The spin gap Δ and the Curie constants of Zn-doped samples, as estimated from an analysis of the magnetic susceptibility, monotonically decrease as the Zn concentration decreases. A nonmagnetic impurity of $S$=1/2 spin ladder system affects the spin gap.

Email: sokubo@kobe-u.ac.jp

Fax: +81-078-803-5654

## I. INTRODUCTION

Low-dimensional spin systems have attracted much attention because they show a variety of magnetic properties [1], such as the spin Peierls transition [2], Haldane and non-Haldane states [3] and a spin liquid state. After the establishment of the Haldane gap system [1], studies of quantum spin systems were extended to complicated spin systems such as the $S$=1/2 spin ladder system $SrCu_2O_3$ [4, 5]. In particular, the effect of a nonmagnetic impurity on a quantum spin chain elucidates the quantum effect as the appearance of magnetic moments with spin correlation around the impurity [6, 7], which is one of typical quantum effects of correlated spins. The nonmagnetic impurity effect in low-dimensional systems with singlet ground states appear as impurity-induced staggered moments with decay distributions, which lead to effective exchange interactions between spins. The effective interactions are mediated by singlet spin pairs and produce exotic ground states. These systems behave as antiferromagnetic finite-length chains or three-dimensional long-range order. In an experimental study, such impurity-induced antiferromagnetic ordering was observed in the impurity-doped spin Peierls system $Cu_{1-x}Zn_xGeO_3$ [8], the Haldane system [9] and the spin ladder systems $Sr(Cu_{1-x}Zn_x)_2O_3$ and $[Ph(NH_3)]([18]crown-6)[Ni(dmit)_2]$ [10, 11]. We focus on the nonmagnetic impurity effect in the $S$=1/2 two-leg spin ladder system because theoretical works have been performed [6, 7]. However, when investigating the magnetic properties, previous works have some problem, such as the obtained sample being a powder and few number of the nonmagnetic impurity concentrations. Therefore, we have tried to synthesize Zn-doped samples of $(pipdH_2)CuBr_4$, chemically called bis(piperidinium)tetrabromocuprate(II), which is one of the most investigated $S$=1/2 spin ladder systems.

The crystal structure of $(pipdH_2)CuBr_4$, which corresponds to $(C_5H_{12}N_2)CuBr_4$ in chemical formula, is monoclinic with space group *p21/c* (No. 14) [12]. The *a*, *b*, and *c* lattice constants are 848.7 pm, 1722.5 pm and 1238.0 pm, respectively. $Cu^{2+}$ ions, which have spin 1/2, form a two-leg ladder along the *a* axis, as shown in Figure 1. The inter-ladder is well separated by piperidinium molecules. The magnetic susceptibility measurements on single crystals under an applied magnetic field of 0.1 T show the general shape of typical low-dimensional magnetic systems. A broad maximum is observed at around 8 K, and exponential temperature dependence is observed below the peak [13]. The Curie-Weiss temperature, $\theta$ = -5.3 K, from a plot of the inverse magnetic susceptibility shows that an antiferromagnetic coupling is dominant in this system. The interchain couplings $J_\perp$ and $J_{//}$ are evaluated to be 13.3 K and 3.8 K, respectively, by using exact diagonalization calculations with 12 spins. Magnon dispersion relations for fully-deuterated (pipdD$_2$)CuBr$_4$ has been reported from the inerastic neutron scattering measurements by Savici *et al*. [14]. They estimated the interchain couplings $J_\perp$, $J_{//}$ and $J_{\text{interladder}}$ to be 1.09 meV (12.6 K), 0.296 meV (3.4 K) and $\leq$ 0.006 meV (0.07 K), respectively, by using the $S$=1/2 two-leg ladder model with a singlet ground-state. From the dispersion relation, (pipdH$_2$)CuBr$_4$ can be consider as an ideal isolated $S$=1/2 two-leg ladder system. Moreover, from $^{14}$N NMR measurements, a one-dimensional Tomonaga-Luttinger liquid (TLL) state, which was predicted theoretically [15, 16], was realized in magnetic fields above the lower critical field $H_{C1}$=6.75 T at which the spin gap closes [17]. The existence of a three-dimensional long-range-order phase below the TLL phase in (pipdH$_2$)CuBr$_4$ is the same as the theoretical prediction. The best way to investigate the effect of a nonmagnetic impurity on the $S$=1/2 two-leg ladder system is to use a well-established $S$=1/2 two-leg ladder system such a (pipdH$_2$)CuBr$_4$. Therefore, we decided to synthesized nonmagnetic-impurity-doped (pipdH$_2$)CuBr$_4$ and to estimate its magnetic properties. In this paper, we report the synthesis of Zn-doped (pipdH$_2$)Cu$_{1-x}$Zn$_x$Br$_4$ samples and estimates of the magnetic impurity and the magnetic properties of the magnetic susceptibility.

## II. EXPERIMENTAL

The sample synthesis followed a procedure similar to Patyal's procedure, except for the mixing ratio of materials [12]. The pure sample ((pipdH$_2$)CuBr$_4$) was prepared by slow evaporation of a solution of piperidinium bromide ((pipdH)Br) (0.004 mol, 0.665 g) and CuBr$_2$ (0.002 mol, 0.447 g) in 23 ml of ethanol. For the impurity-substituted sample, ((pipdH)Br), ZnBr$_2$ and CuBr$_2$ were mixed in the stoichiometric proportion of (pipdH)$_2$Cu$_{1-x}$Zn$_x$Br$_4$. After the starting materials had been mixed in ethanol at room temperature, a methanol solution was put into a beaker covered with a laboratory film with small holes for slow evaporation of the solvent from the methanol solution. To crystallize the sample, we left it under a fixed temperature at 303 K. After 1-2 weeks, small, shiny, black crystals of (pipdH)$_2$Cu$_{1-x}$Zn$_x$Br$_4$ were formed. Five different Zn-concentration samples, x=0, 0.01, 0.02, 0.05, and 0.1, were obtained. A typical crystal is shown in Figure 2.

Powder X-ray diffraction (XRD) measurements to examine the samples were performed by using "Rigaku MiniFkex II" equipment with a Cu X-ray source. The magnetic susceptibility measurements were performed under a magnetic field of 0.1 T by using a superconducting quantum interference device (SQUID) magnetometer (Quantum design MPMS) in the temperature range from 1.8 K to 300 K. The samples were used in powder form for comparison with previous works.

## III RESULTS AND DISCUSSION

Figure 3 shows powder X-ray diffraction patterns of (pipdH)$_2$Cu$_{1-x}$Zn$_x$Br$_4$ (x=0, 0.01, 0.02, 0.05, 0.1) and the calculated value. The crystal parameter of (pipdH$_2$)CuBr$_4$ from Patyal *et al.*'s report [12] is used for the calculated value pattern. The XRD peaks of (pipdH)$_2$Cu$_{1-x}$Zn$_x$Br$_4$ are in good agreement with the calculated value, which suggests that the overall crystal structure of the Zn-doped samples does not change from that of the pure sample of (pipdH$_2$)CuBr$_4$. However, linewidths of the peaks increase slightly as the Zn concentration is increased, which may be caused by the crystal structure of Zn-doped samples having a discontinuous deformation due to the different ionic radii of Zn$^{2+}$ and Cu$^{2+}$ ions because impurity Zn$^{2+}$ ions are substituted for Cu$^{2+}$ ions from the charge neutrality requirement.

Figure 4 shows the magnetic susceptibility $\chi$, $\chi T$ and the inverse magnetic susceptibility $1/\chi$ as functions of the temperature for a pure sample of (pipdH$_2$)CuBr$_4$. Raw data on the magnetic susceptibility, as shown in Figure 4(a), clearly show no Curie tail at low temperatures, which means that the sample is very clean and has a low impurity concentration. A large diamagnetic component is expected in (pipdH$_2$)CuBr$_4$ because an organic material contains relatively large numbers of nonmagnetic ions, which have the core diamagnetisms of the atoms. The temperature-independent diamagnetic component, $\chi_0$, is estimated to be -2.22 x 10$^{-4}$ emu/mol from the negative slope of the $\chi$T-T plot at high temperatures, which should be a certain constant value in the paramagnetic state. Here, the g-value is used as 2.00. On the other hand, the core diamagnetism for (pipdH$_2$)CuBr$_4$ is estimated to be -2.52 x 10$^{-4}$ emu/mol by using Pascal's constants from the literature values for Cu$^{2+}$, C$^{2-}$, H$^{+}$, N$^{3-}$ and Br$^{-}$. The estimated diamagnetic component $\chi_0$ coincides with the calculated core diamagnetism. The Curie-Weiss temperature $\theta$ and the Curie constant $C$ are estimated to be -3.63 K and 0.411 emu K/mol, respectively, from a fit to $\chi = \chi_0 + C/(T-\theta)$ for 50 K < $T$ < 300 K. Those are in good agreement with previous reports [13].

The singlet-triplet energy gap $\Delta$ is estimated to be 8.1 K from Troyer's equation $\chi \propto T^{-1/2} \exp(-\Delta/T)$ [18] by fitting the magnetic susceptibility below the peak. The estimated gap $\Delta$ = 8.1 K consists with obtained gap from the high-field magnetization measurement (gap $\Delta$ = 9.5 K, $H_{C1}$ = 6.6 T) [13]. As mentioned above, the procedure for the diamagnetic component correction can be considered to be correct, and our sample quality is the same as those in previous works. The diamagnetic correction was performed for Zn-doped samples of (pipdH)$_2$Cu$_{1-x}$Zn$_x$Br$_4$ (x=0.01, 0.02, 0.05, 0.10) by using the above-mentioned procedure. The obtained diamagnetic components $\chi_0$, which are good agreement with the core diamagnetism from Pascal's law, are shown in Table 1. Figure 5 shows the magnetic susceptibilities of (pipdH)$_2$Cu$_{1-x}$Zn$_x$Br$_4$ (x=0.01, 0.02, 0.05, 0.10), from which the diamagnetic components $\chi_0$ has been subtracted. Each magnetic susceptibility curve is shifted by 0.005 emu/mol for clarity. Although the magnetic susceptibility at low temperatures increases as the Zn-doped concentration is increased, even at a 10% Zn-doping concentration, only a few data points show a Curie tail. We cannot estimate the Curie term for the impurity from this Curie tail because the tail is small. The peak temperatures of the magnetic susceptibilities slightly decrease as the Zn-doping concentration is increased. The estimated Curie-Weiss temperatures $\theta$ and the Curie constants $C$ from plots of the inverse magnetic susceptibility in the temperature range from 50 K to 300 K are shown in Table 1.

In the analysis of the gap $\Delta$, a large discrepancy between the magnetic susceptibility curves and the fitting curves of Troyer's equation occurs at low temperatures for high Zn concentration of 10% and 5%. A possible cause of this discrepancy is that the Zn-doped sample at a high concentration does not contain the Zn concentration of the mixing ratio. The actual Zn concentration of the doped samples can be considered to be 1 or 2% smaller than the mixing ratio. The Zn concentrations from the best fits to Troyer's equation are 4% and 7% for Zn-5% and Zn-10% mixing ratios, respectively, as shown by percents in parentheses in Table 1. We use these Zn concentrations as the actual concentrations to estimate the Curie constants for high-concentration Zn-5% and Zn-10% samples. For the Zn-2% sample, we use a mixing ratio 2% because there is no large discrepancy between the magnetic susceptibility curves and the fitting curves for Troyer's equation. The Curie constants $C$ show a tendency to decrease as the Zn-doping concentration is increased. On the other hand, the Weiss temperatures $\theta$ are almost constant, which means that the dominant coupling does not change and that the number of effective spins is decreased by Zn doping. Moreover, the spin gap $\Delta$ monotonically decrease as the Zn concentration is decreased. For further investigation of the magnetic properties, high-field magnetization and microscopic measurements are required.

### III. CONCLUSION

To examine the nonmagnetic impurity effect of a $S=1/2$ spin ladder system, we succeeded in synthesizing Zn-doped (pipdH)$_2$Cu$_{1-x}$Zn$_x$Br$_4$ (x= 0, 0.01, 0.02, 0.05 0.1) samples for the first time. The Curie constant, the Weiss temperature and the spin gap $\Delta$ of Zn-doped (pipdH)$_2$Cu$_{1-x}$Zn$_x$Br$_4$ have been estimated systematically. The spin gap $\Delta$ and the Curie constant of Zn-doped samples monotonically decrease as the Zn concentration decreases.

ACKNOWLEDGMENTS

This work was partly supported by a Grant-in-Aid for Creative Scientific Research (No. 19GS1209) from the Japan Society for the Promotion of Science. One of the authors (CY) thank developer of VESTA, help us to visualize crystal structure and to understand relation between ions [19] and is also grateful for the financial support from the Molecular Photoscience Research Center of Kobe University.

Figure Captions

Fig. 1. Crystal structure of (pipdH)$_2$CuBr$_4$. The spin ladder lies along the *a*-axis.

Fig. 2. Close-up photo of a (pipdH)$_2$Cu$_{0.9}$Zn$_{0.1}$Br$_4$ crystal. The dimensions of the largest sample are about 7 x 15 x 0.5 mm$^3$.

Fig. 3. Powder X-ray diffraction patterns of (pipdH)$_2$Cu$_{1-x}$Zn$_x$Br$_4$ (x=0, 0.01, 0.02, 0.05, 0.1) and the calculated value.

Fig. 4. Magnetic susceptibility of (pipdH)$_2$CuBr$_4$. (a) Magnetic susceptibility of raw data (before subtraction of a diamagnetic term $\chi_0$) as a function of temperature. The inset figure is an enlarged plot for low temperature. (b) $\chi$T plots after subtraction of a diamagnetic term $\chi_0$. (c) Inverse magnetic susceptibility after subtraction of a diamagnetic term $\chi_0$.

Fig. 5. Temperature dependence of the magnetic susceptibility of (pipdH)$_2$Cu$_{1-x}$Zn$_x$Br$_4$ (x=0, 0.01, 0.02, 0.05, 0.1).

Table 1. Obtained parameters and comparison with previous reports.

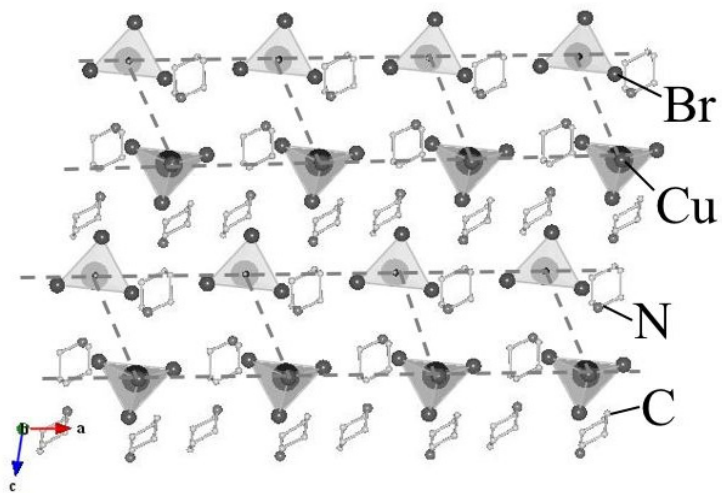

Fig. 1. Crystal structure of (pipdH)$_2$CuBr$_4$. The spin ladder lies along the *a*-axis.

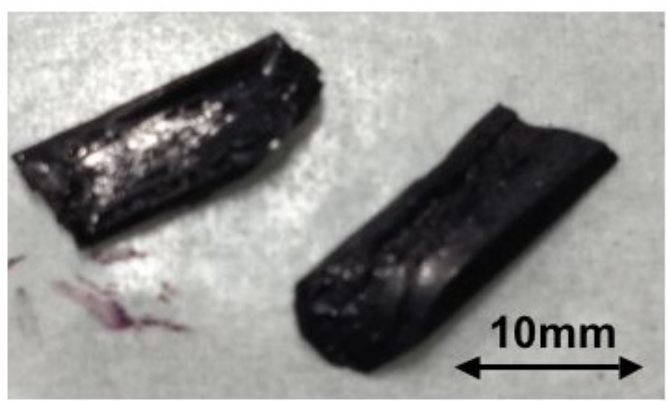

.

Fig. 2. Close-up photo of a (pipdH)$_2$Cu$_{0.9}$Zn$_{0.1}$Br$_4$ crystal. The dimensions of the largest sample are about 7 x 15 x 0.5 mm$^3$.

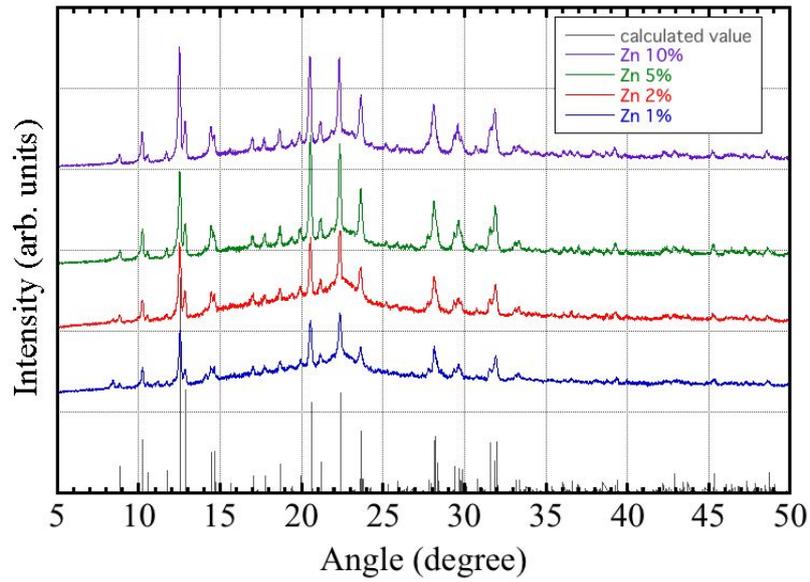

Fig. 3. Powder x-ray diffraction patterns of (pipdH)$_2$Cu$_{1-x}$Zn$_x$Br$_4$ (x=0, 0.01, 0.02, 0.05, 0.1) and the calculated value.

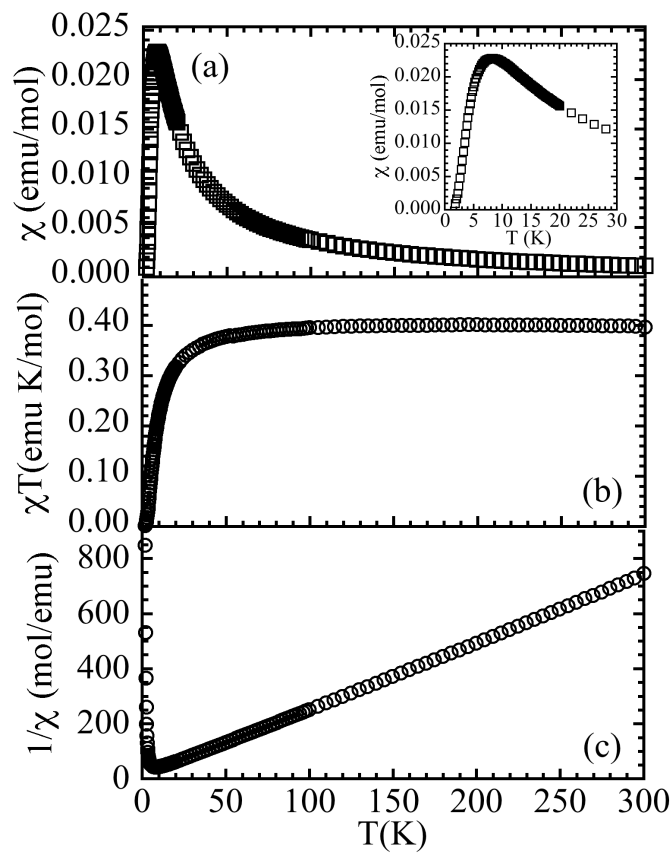

Fig. 4. Magnetic susceptibility of (pipdH)$_2$CuBr$_4$. (a) Magnetic susceptibility of raw data (before subtraction of a diamagnetic term $\chi_0$) as a function of temperature. The inset figure is an enlarged

plot for low temperature. (b) χT plots after subtraction of a diamagnetic term $\chi_0$. (c) Inverse magnetic susceptibility after subtraction of a diamagnetic term $\chi_0$.

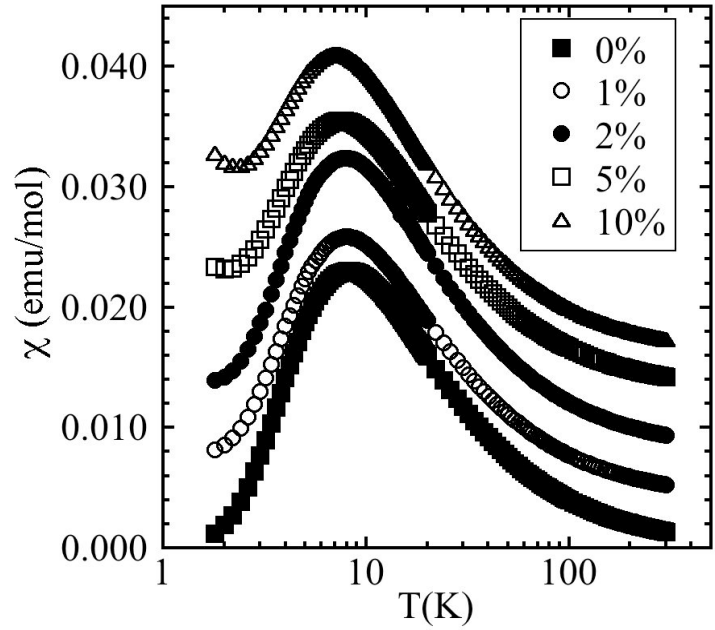

Fig. 5. Temperature dependence of the magnetic susceptibility of $(pipdH)_2Cu_{1-x}Zn_xBr_4$ (x=0, 0.01, 0.02, 0.05, 0.1).

Table 1  Obtained parameters and comparison with previous reports.

|  | C (emu K/mol) | $\theta$ (K) | $\Delta$ (k) | $\chi_0$ (x$10^{-4}$ emu/mol) | $\chi_0$ (x$10^{-4}$ emu/mol) Pascal's law |
|---|---|---|---|---|---|
| Zn 0% (Watson [11]) | 0.433 | -5.3 | 9.5 |  |  |
| Zn 0% | 0.411 | -3.63 | 8.1 | -2.22 | -2.52 |
| Zn 1% | 0.378 | -3.17 | 7.39 | -2.30 | -2.53 |
| Zn 2% | 0.407 | -1.81 | 7.74 | -2.47 | -2.54 |
| Zn 5% (4%) | 0.354 | -2.09 | 7.24 | -2.48 | -2.57 |
| Zn 10% (7%) | 0.376 | -3.45 | 7.16 | -2.43 | -2.62 |